ARTÍCULO ORIGINAL

# El egresado de la carrera Ciencias de la Información y su inserción en la gestión de mercadotecnia

Incorporation of information science graduates to marketing management tasks

Lic. Carlos Luis González-Valiente,[I] Dra. C. Magda León Santos,[II] Dra. C. Zoia Rivera[II]

[I] Departamento de Informática y Gestión de la Información del Grupo Empresarial de la Industria Sidero Mecánica (GESIME). La Habana, Cuba.
[II] Facultad de Comunicación. Universidad de La Habana. La Habana, Cuba.

**RESUMEN**

En el estudio se busca identificar las posibilidades que ofrece la formación académica en Cuba para el desempeño del profesional de la información como gestor de mercadotecnia. Se abordan teóricamente las funciones básicas del mercadeo y su dimensión informacional, las particularidades del trabajo del gestor de mercadotecnia y las competencias que funda el Plan de Estudios D sobre los egresados de la carrera Ciencias de la Información en el contexto cubano. A partir del análisis de contenido del plan de estudio y las entrevistas aplicadas a diez profesionales de la información que fungen como gestores de mercadotecnia, se visualizaron algunas competencias indispensables que deben formarse en la carrera y consolidarse, posteriormente, en el ejercicio práctico.

**Palabras clave:** mercadotecnia, gestor de mercadotecnia, profesional de la información, competencias, plan de estudios, Cuba.

**ABSTRACT**

The study aims to identify the possibilities offered by academic studies in Cuba for the training of information professionals as marketing managers. A theoretical analysis is conducted of the basic functions of marketing and its information dimension, the specific features of the work of the marketing manager, and the competencies developed by Curriculum D in information science graduates in the Cuban context. Based on the analysis of curricular contents and interviews with ten information professionals working as marketing managers, determination was made of some indispensable competencies which should be developed during the training and later on during service practice.

**Key words:** marketing, marketing manager, information professional, competencies, curriculum, Cuba.

## INTRODUCCIÓN

La información es un elemento que se ha convertido en factor de cambio y transformación para el cuerpo teórico y práctico de la disciplina Mercadotecnia.[1-2] Publicaciones de esta área se han orientado a explorar cómo los factores informacionales pueden afectar el conocimiento de la organización,[3] lo cual está básicamente mediado por la manera en que las personas interpretan la información del mercado. *Day* afirma que tales interpretaciones son facilitadas por los modelos mentales de los decisores;[4] y los resultados de tal proceso, a instancias de *Daft* y *Weick*, tienen grandes implicaciones sobre las investigaciones y las prácticas gerenciales.[5] La indagación relativa a estos fenómenos emerge a partir del comportamiento informacional que los gestores de mercadotecnia asumen para concebir y dar respuesta a sus tareas de trabajo. En opinión de *Fleisher*, *Wright* y *Allard* existen cuatro técnicas de gestión de información que estos especialistas deben integrar para la planificación y ejecución de sus labores: inteligencia competitiva, gestión de relación de clientes, minería de datos e investigación de mercados.[6]

La plataforma informacional sobre la cual la actividad mercadotécnica está soportada la convierte en un escenario proclive para la investigación desde el dominio de las Ciencias de la Información. *Holland* y *Naudé* afirman que esta es una de las áreas que incide en actividades genéricas como: el análisis de mercado y recolección de datos; el análisis de consumidores, segmentación y orientación; los productos, estrategias y objetivos de mercadotecnia; la comunicación con socios comerciales y la implementación de lo planificado.[7]

Desde una perspectiva inversa, la incidencia de la mercadotecnia sobre las ciencias de la información se ha dado en la manera en que los elementos de mercadeo pueden ser aplicados a los productos y servicios de información que se generan en las organizaciones.[8-11] Esto ha presupuesto que el profesional de la información asuma una actitud empresarial a la hora de ejecutar sus ofertas. Es por eso que las nociones teórico-prácticas relativas al mercadeo han llegado a ser una temática curricular de necesario interés para el campo de las Ciencias de la Información.[9] Los conocimientos obtenidos han propiciado una mayor inserción en un mercado de trabajo más amplio, ya que la confluencia de las competencias informacionales sobre las mercadotécnicas le permiten desempeñarse como especialistas en información de mercadotecnia, perfil que *Cronin*, *Stiffler* y *Day* ubican dentro de un gran nicho de ocupaciones en el cual coexisten desde analistas de sistemas de información hasta especialistas de comunicaciones.[12]

De ahí que, el abordaje acerca del binomio *Mercadotecnia-Ciencias de la Información* se hará desde la perspectiva en la que el modelo formativo del profesional de la información en el contexto cubano provee un marco sólido de competencias para incidir en esta actividad. El objetivo general de este estudio está encaminado a determinar las posibilidades que tienen tales profesionales como gestores de mercadotecnia, a partir de los contenidos y habilidades que se forman durante el proceso de estudio de la carrera. Para esto, se abordan algunos aspectos teóricos asociados al mercadeo, en donde se pueda caracterizar a este desde una perspectiva informacional. También serán esclarecidas las particularidades del trabajo del gestor de mercadotecnia y, además, se compararán qué conocimientos y habilidades formados en la carrera se corresponden con la realidad de este tipo de trabajo.

MERCADOTECNIA: ELEMENTOS BÁSICOS

La Asociación Americana de Mercadotecnia (AMA, por sus siglas en inglés) definió a la mercadotecnia como "la actividad, conjunto de instituciones y procesos para crear, comunicar, entregar e intercambiar ofertas que tienen valor para los clientes, socios y la sociedad en general".[13] En esta definición subyace su carácter estratégico; de ahí que *Keegan* y *Rolf* la definen como "el proceso de centrar los recursos y los objetivos de una organización en las oportunidades y necesidades de su entorno".[14] Su ejecución tiene lugar por la confluencia de cuatro factores clave. Estos son:

1. Dos o más partes (individuos u organizaciones) con necesidades insatisfechas.

2. El deseo y la habilidad de estas partes para satisfacerlas.

3. El modo en que esas partes se comunican entre sí.

4. Algo para intercambiar.[15]

Estos elementos constituyen el punto de partida para que la mercadotecnia sea concebida como el proceso administrativo que se interesa por las relaciones de consumo, la cual sustentada en el principio de intercambio incide sobre la coexistencia de tres categorías genéricas: el individuo, la organización y el mercado. El individuo, como aquel objeto humano que asume una actitud activa o pasiva sobre el consumo del producto/servicio. La organización, como la diseñadora y ejecutora de las estrategias y acciones de mercadeo. Y, por último, el mercado, como escenario donde el consumo se materializa, y en donde se establece, además, el nexo entre la organización y el individuo.

Para *Kotler*, el proceso mercadotécnico comprende una secuencia de cuatro actividades genéricas fundamentales: 1) el análisis de las oportunidades, 2) el diseño de estrategias, 3) la planificación de programas y 4) la organización y el control del esfuerzo de mercadotecnia.[16] Esta misma idea es tomada en cuenta por *Kurowski* y *Sussman,* quienes consideran que lo primero es conocer ampliamente el mercado, luego definir un método para informarle al consumidor el propósito determinado y, por último, implementar lo definido.[17] A continuación se describen cada una de estas fases principales:

- La investigación de mercados.

    Esta tarea investigativa es definida por *Kotler* como el "diseño sistemático, recolección, análisis y presentación de la información y descubrimientos relevantes acerca de un sistema de mercadotecnia específica a la que se

enfrenta la empresa".[16] Su resultado provee de la información necesaria para la toma de decisiones de corto y largo plazo en las empresas.[18] Sobre el análisis de los elementos informacionales, la mercadotecnia dispone de métodos que son generalmente de tipo estadísticos. Las teorías de la información que lo avalan proponen metodologías y modelos particulares que permiten desarrollar las interpretaciones apropiadas.[19,20]

Tal investigación no es siempre aplicada, ya que ella opera bajo la definición de un problema específico, por lo que las organizaciones pueden hacer uso del constante análisis del entorno, el cual según *Buttery* y *Tamaschke*, "se refiere a la adquisición de información sobre eventos y tendencias en el entorno de una organización, y del conocimiento que sería de ayuda para los altos ejecutivos en la identificación y comprensión de las oportunidades estratégicas y las amenazas".[21]

- La estrategia de mercadotecnia.

La AMA declara que el diseño de la estrategia implica la toma de decisiones sobre cada uno de los componentes mercadotécnicos.[13] Según *Kurowski* y *Sussman*, "la información de la investigación y análisis de mercado es la base para la planificación estratégica; y la aplicación de los instrumentos de mercadotecnia, para alcanzar los objetivos de ventas y la posición deseada en relación con los productos y grupos destinatarios".[17] Esto contribuye a conocer las debilidades, amenazas, fortalezas y oportunidades (DAFO) de las organizaciones, lo cual constituye una de las más importantes herramientas estratégico- informativas.[18] Los componentes que integran tal estrategia son identificados y comúnmente conocidos como las cinco P: producto, precio, plaza (canales de distribución), promoción y personas. Del trabajo con ellos se deriva una actividad conocida como mezcla de mercadotecnia, la cual busca la continua combinación de estos.[17]

- La organización, control y el plan de mercadotecnia.

La organización de mercadotecnia es la manera en la que el departamento en cuestión se relaciona con otras funciones de la empresa y define cómo deben ser articuladas y realizadas las estrategias,[16] mientras que los elementos del control, de acuerdo con *Kotler*, pueden hacerse sobre el plan anual, la rentabilidad, la eficiencia o lo estratégico. Para este autor, en todas ellas coexiste un elemento básico: la información, la cual permite develar el estado real del comportamiento de cada uno de estos parámetros, de manera que facilite medir el esfuerzo acumulado.[16]

Por otro lado, el plan se encarga de reflejar todo lo que la estrategia ha tomado en cuenta para su diseño. *Kurowski* y *Sussman* afirman que este se realiza a partir de todo el trabajo informacional desarrollado en cada actividad anterior. Además, esto debe guardar una relación directa con los recursos, actividades, objetivos y propósitos de la organización. Tales autores conciben que el plan debe presentar elementos clave de publicidad y promoción, distribución, organización mercadotécnica e instrumentos y presupuesto.[17]

**Dimensión informacional de la mercadotecnia**

Hasta ahora se ha estado destacando cómo cada una de las actividades genéricas que comprende un proyecto depende de la información. Con el propósito de consolidar fundamentos al respecto, en este epígrafe se profundiza en el carácter informacional de este ejercicio, No precisamente orientándose hacia una definición conceptual, sino reflejando, mediante un análisis de la literatura, la manera en la que esta área de estudio asume el fenómeno informativo.

Para la mercadotecnia, la información es asumida como un evento que tiende a cambiar relaciones entre sus productores (gestores de mercadotecnia) y los consumidores de esta en el mercado.[22] De forma amplia, esta área del conocimiento ha abordado tres aspectos fundamentales asociados a la información: 1) los procesos informacionales y la definición, alcance e implicación de estos para las organizaciones, 2) los objetos y estructuras presentes en el contexto de mercado y 3) el comportamiento informacional de las personas (organizaciones/consumidores) que la manipulan. La AMA no destina una concepción para este término, sino que lo define como parte de categorías que responden a procesos informacionales en sí, tales como: control, procesamiento, búsqueda y clasificación de información, así como publicidad y nivel informacional.[13]

El interés de esta disciplina por estudiar cuestiones asociadas a la información deviene por la gran capacidad que esta posee para evidenciar las situaciones del mercado,[23] divisar eventos futuros,[24] proveer ventaja competitiva[1-25] y dotar de experiencia y aprendizaje a la organización.[26] Tales condiciones contribuyen a que la información sea enfocada por la alta dirección de las organizaciones como elemento estratégico para el desarrollo de sus funciones. De ahí, que hace falta aumentar las inversiones en la gestión de los conocimientos con vistas a actuar según los modelos actualizados, ya que "las decisiones necesarias resultan de la interacción del conocimiento y la información a nivel gerencial".[14]

Sobre este aspecto, el trabajo más integral es el de la autora *Christine Moorman*, quien definió claramente los procesos organizacionales de información de mercado. Su aporte radica en la enmarcación del uso aislado de este elemento en cuatro procesos clave, los cuales son expuestos a continuación:[27]

1. *Procesos de adquisición de información*: se basan en la recolección primaria o secundaria de información proveniente del entorno externo de la organización.

2. *Procesos de transmisión de información*: proceso en el cual la información es difundida entre los usuarios dentro de las organizaciones.

3. *Procesos de utilización conceptual*: la manera en la cual las organizaciones procesan la información y a su vez se comprometen con ella. Estos constan de un subproceso 1, el *compromiso hacia la información*, a través del cual la organización reconoce el valor de la información para tomar decisiones. Y un subproceso 2, el *procesamiento de la información*, donde se le atribuye significado a la información utilizada.

4. *Procesos de utilización instrumental*: uso de la información de mercado para influenciar la estrategia de mercadotecnia en relación con las acciones. Aquí son incluidos tres subprocesos que inciden sobre el uso de la información en la (1) *toma*, (2) *implementación* y (3) *evaluación* de las decisiones.

Cada uno de estos procesos es articulado por sistemas de información, como los que propone *Kotler*: Sistema de Información de Mercadotecnia, Sistema de Inteligencia de Mercadotecnia, Sistema de Investigación de Mercadotecnia y Sistema de Apoyo a las Decisiones.[16] Toda la base conceptual de los procesos mencionados también ha servido de referente para el desarrollo de múltiples estudios empíricos que se han proyectado en examinar la manera en la que los gestores de mercadotecnia usan la información y sus fuentes (Ej.; *Ashill* y *Jobber*; *Bennet*; *Cillo*, *De Luca* y *Troilo* y *Du*).[28-31]

**El gestor de mercadotecnia**

La AMA aporta una concepción sobre estos profesionales, pero desde el carácter de sus funciones gerenciales, comerciales o departamentales, las cuales caen bajo una categoría que responde a la responsabilidad organizacional. Ellos son definidos como los designados para ejercer funciones de: investigación de mercados, planificación de productos, fijación de precios, mezcla de promoción, servicio al cliente y otras.[13] *Borden* asevera que este gestor busca la integración de los ingredientes y fórmulas de las operaciones a través de las cuales diseña estrategias.[32] Según enuncian *Brownlie* y *Saren*, el conocimiento y sabiduría de estos debe estar encaminado a "identificar las necesidades y deseos insatisfechos; definir y medir su magnitud; determinar qué mercado objetivo la organización puede atender mejor; decidir sobre los productos, servicios y programas apropiados para servir a estos mercados, y llamar a todos en la organización para pensar y servir al cliente".[33]

Las ideas que se han abordado alrededor de la gestión mercadotécnica son las que configuran el sistema de competencias de estos profesionales. Ellos se enfrentan a constantes desafíos debido al entorno tan cambiante que caracteriza a las organizaciones, donde no solo tienen que lidiar con los grupos internos, sino también con disímiles contingencias externas.[34-36] Para esto, hacen un uso intensivo de información, ya que esta les permite identificar las necesidades del mercado, coordinar eficientemente sus funciones con las del resto de los actores que integran la organización y comunicar los procederes de mercadotecnia.[37]

**El profesional de la información en Cuba**

La comisión nacional de la carrera de Ciencias de la Información, de Cuba, es quien instituye las competencias medulares en sus egresados sobre la base de un conocimiento científico. Tal comisión declara que los profesionales de la información son aquellos que "pueden desempeñarse en cualquier organización que genere, posea, acceda y utilice información de forma intensiva",[38] aunque aportar una concepción más o menos precisa sobre un profesional puede tornarse algo complejo por las cambiantes relaciones interdisciplinares y de conocimientos de los programas de estudio. Por esto, para definir mejor a este gremio se expondrá su sistema básico de competencias a partir de lo que su formación concibe.

El sistema y alcance de las competencias está básicamente integrado por tres categorías particulares: los conocimientos, las habilidades y los atributos personales. En el contexto de la enseñanza superior en Cuba, estas son moldeadas por el "Plan de estudios D",[38] el cual esboza todo el contenido curricular de la especialidad. A este plan lo han antecedido otros cuatro, el primero de los cuales, el Plan "A", entró en vigor en el año 1977. Es importante destacar que, a nivel de asignaturas, los contenidos académicos comprendidos en estos planes se actualizaban constantemente. Con eso se contribuyó a la flexibilización de la enseñanza y a la sustitución de conocimientos que iban perdiendo la importancia para el desarrollo teórico, práctico y social de la especialidad. A continuación se lista el conjunto de competencias contempladas en el Plan:[38]

**Competencias profesionales**

1. Conocimientos.

   a) Fundamentos de la profesión informativa.

   b) Tratamiento de la información.

   c) Diseño e implementación de sistemas, productos y servicios de información.

   d) Alfabetización y socialización de la información.

   e) Gerencia de la información y del conocimiento.

   f) Investigación.

   g) Pedagogía y docencia.

   h) Tecnologías de la información.

2. Habilidades.

   a) Manejo de fuentes de información.

   b) Formación y capacitación en el uso de la información.

   c) Comunicación.

   d) Toma de decisiones y solución de problemas basados en información.

   e) Trabajo en equipo.

   f) Gestión de documentos, de información y del conocimiento.

   g) Almacenamiento, conservación y preservación de registros de información y conocimientos.

   h) Diseño, implementación y evaluación de productos, servicios y sistemas de información.

   i) Dominio de los procesos asociados a la información.

3. Atributos personales.

   a) Ética profesional.

   b) Cultura informacional.

   c) Motivación y vocación hacia el ejercicio profesional.

   d) Calidad en el ejercicio de la profesión.

   e) Compromiso social.

De los conocimientos, habilidades y atributos señalados, para el desempeño del profesional de la información como gestor de mercadotecnia, parecen ser de mayor importancia los que se orientan a la gestión, la investigación y la producción de servicios y productos de información. Entre estas competencias se pueden mencionar la gestión de documentos, de información y de conocimientos; el diseño, implementación y evaluación de productos, servicios y sistemas de información; la metodología de la investigación; la toma de decisiones y la solución de problemas basados en información; entre otras.

En el estudio se busca identificar las posibilidades que ofrece la formación académica en Cuba para el desempeño del profesional de la información como gestor de mercadotecnia.

## MÉTODOS

Se realizó un estudio descriptivo, en el cual se recurrió al análisis documental clásico para recopilar información en torno al objeto de estudio. A su vez, el uso de la técnica de análisis de contenido, aplicado al plan de estudios "D", proporcionó los elementos necesarios para advertir las dimensiones previstas en la formación académica del profesional de la información.

Con vistas a comparar lo advertido con la realidad de trabajo en el ámbito de mercadotecnia, se realizaron las *en*trevistas a diez profesionales de la información que se desempeñan en estas funciones. Las entrevistas se centraron en los tópicos de interés para comprender las vivencias y opiniones de los informantes respecto a las competencias que ellos consideran necesarias para desarrollar las actividades de mercadotecnia. Es de señalar que para las entrevistas no se hizo la selección de una muestra, sino que se contactó con todos los egresados de la carrera que se desempeñan como gestores de mercadotecnia. En La Habana, en total, se pudo identificar a diez especialistas, cifra bastante reducida, pero que, a la vez, potencia la necesidad del presente estudio. Por otro lado, ninguno de los informantes hizo la carrera soportada en el Plan "D", que comenzó en el curso académico 2008-2009; por tanto, la información obtenida a partir de las entrevistas sirvió para comparar las competencias indispensables para el desempeño en la mercadotecnia con lo previsto en el plan de estudios.

El modo de aplicación de las entrevistas fue personal y directo en cada uno de los ambientes de trabajo correspondientes. La duración fue aproximadamente de una hora, y se cumplió en cada caso con los objetivos propuestos. En la tabla 1 se presenta el entorno laboral en que se desempeñaron los entrevistados.

**Tabla.** Caracterización general del ámbito laboral de los entrevistados

| Sector industrial | Entrevistados | |
|---|---|---|
| | Cantidad | Porcentaje (%) |
| Comercio de importación | 3 | 30 |
| Comercio Interior | 1 | 10 |
| Cultura y Arte | 1 | 10 |
| Ciencia y Tecnología | 2 | 20 |
| Turismo | 2 | 20 |
| Salud | 1 | 10 |
| Total | 10 | 100 |
| Departamento | | |
| Mercadotecnia | 8 | 80 |
| Otros Departamentos | 2 | 20 |
| Total | 10 | 100 |
| Posición | | |
| Jefe de Departamento | 2 | 20 |
| Vicedirector | 1 | 10 |
| Especialista principal | 6 | 60 |
| Total | 100 | 100 |

Además de las cuatro actividades genéricas de mercadotecnia (investigación de mercados/análisis del entorno, diseño de la estrategia/mezcla de mercadotecnia, organización y control, y plan de mercadotecnia), desde una perspectiva informacional, la guía de la entrevista contó con la pregunta orientada hacia la obtención de la información en torno a la contribución de los contenidos de la carrera Bibliotecología y Ciencia de la Información al desempeño profesional en mercadotecnia.

## RESULTADOS

En la selección escogida hubo una mayor representatividad de mujeres (7) que de hombres (3), y alcanzó en su conjunto una edad promedio de 40 años. Todos se habían graduado del nivel superior y contaban con estudios de posgrado, entre los que figuraron 3 másteres y 1 doctora; el resto poseía título de licenciado (6). El promedio de años como egresados de la carrera fue de 11, de los cuales habían dedicado 8 años como promedio al ejercicio de mercadotecnia.

De las actividades genéricas expuestas en la entrevista, 7 de los especialistas aseveraron realizar la primera de ellas: el análisis del entorno e investigaciones de mercados (Fig. 1). En ella, todos (10) identificaron, recolectaron y evaluaron las fuentes de información. Uno de los entrevistados aseguró: "mi habilidad con la información ha contribuido a la identificación y validación de sitios web especializados para aplicar encuestas que tributan a una profunda investigación de mercados".

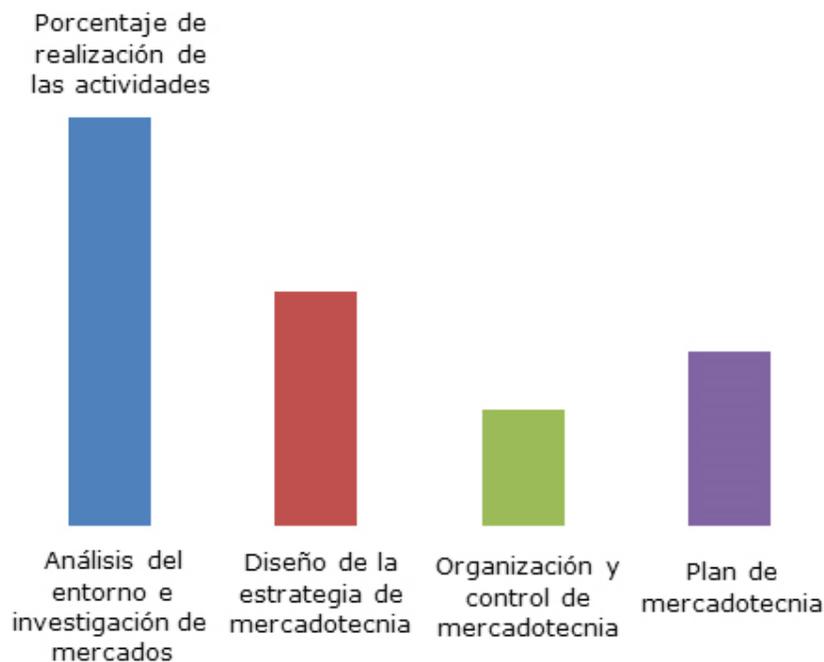

**Fig. 1.** Actividades genéricas de mercadotecnia que realizan los entrevistados.

Solo dos organizaciones contaron con un sistema de información de mercadotecnia institucionalizado. Al respecto, un especialista opinó: "el sistema que uso es el resultado de un trabajo de investigación que realicé, el cual incluyó, además del diseño, su evaluación". El resto de los especialistas reemplazaron las funciones de estos sistemas por otras herramientas tecnológicas, que les permitían procesar datos e información. Entre estas herramientas se destacó el uso de Microsoft Excel.

De los entrevistados, 6 participaron activamente en el diseño e implementación de investigaciones para el análisis del mercado. La herramienta más utilizada para esto fue la encuesta, aunque, en otro sentido, otro entrevistado testificó: "aplico entrevistas personales directas porque me permiten adquirir información circunstancial que me ayuda a distinguir patrones y comportamientos en los clientes durante su acto de consumo".

Los 10 especialistas operaban en el análisis de información de mercado, quienes declararon que se apoyaban en modelos mentales y cognitivos que les habían sido estructurados a partir de lo aprendido en la carrera y del posterior ejercicio profesional. Se develó una fuerte tendencia al uso de métodos y herramientas estadísticas para el procesamiento de la información. En este sentido, una de las informantes subrayó: "para el procesamiento de la información, los métodos estadísticos me ayudan a hacer análisis contables; aunque también mis conocimientos sobre los procesos informacionales me permiten identificar tendencias que a veces no son visibles y eso es clave para la investigación de mercados".

De los especialistas, 8 comunicaban a las instancias pertinentes los resultados de la investigación o el análisis del entorno. Al respecto, 2 entrevistados puntualizaron que técnicas documentales, como el resumen, contribuyen a que los informes de grandes volúmenes puedan ser reducidos a pocas cuartillas, factor que constituye una exigencia de consulta por parte de los ejecutivos de otros niveles. Relativo a la información resultante de la investigación preliminar, uno de los entrevistados afirmó: "es la que me permite diseñar estrategias competitivas y ver en qué medida se está evolucionando o no en materia de mercado".

La actividad de gestión de información resultó ser la de mayor relevancia (Fig. 2). Uno de los informantes destacó: "la gestión de información de productos me facilita definir y diseñar herramientas e indicadores de calidad para interpretar el estado de satisfacción de los clientes".

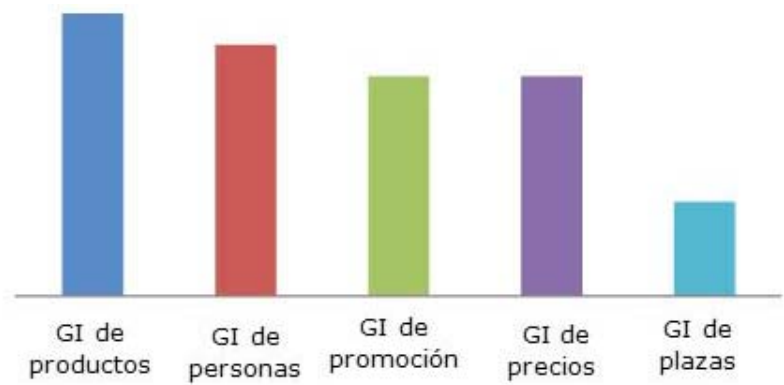

**Fig. 2.** Incidencia de la actividad de gestión de información (GI) sobre los componentes de mercadotecnia.

De los entrevistados, 7 colaboraban directamente en la estrategia de promoción, mientras que 4 participan en el diseño de campañas promocionales y de comunicación. Todos consideraron que la gestión de información les ayudaba a definir y a concebir elementos comunes que no debían faltar en el desarrollo del proyecto. Una entrevistada ratificó: "gestiono mucha información para la promoción, específicamente fotos y textos para los catálogos y puntos de venta. Esa actividad contribuye a que posteriormente se determine a qué polo turístico se priorizará para ejecutar en este las acciones de promoción".

De los especialistas, 7 aseguraron haber diseñado productos y servicios como parte de nuevas ofertas para el mercado. De ellos, 3 alegaron que estos no habían sido de información, pero estaban basados en ella para su concepción; mientras que 4 afirmaron haber diseñado productos y servicios informativos, entre los que habían figurado, indistintamente, boletines impresos y electrónicos, catálogos de productos, servicio de alertas y nuevas notificaciones, servicio de transferencia de información a usuarios y clientes, sistemas de quejas y sugerencias y base de datos para el trabajo interno. Otros 4 habían colaborado en el diseño de intranet y sitios web, específicamente con lo relacionado a la arquitectura de información y la definición de contenidos y elementos de usabilidad.

Respecto al trabajo con el componente "personas", una de las especialistas expuso: "los estudios de usuarios y clientes me ayudan a percibir tendencias de satisfacción y fidelidad en estos, lo cual constituye un elemento primordial para el diseño de estrategias de comunicación, así como de diseño de nuevos productos y servicios". Solo 3 entrevistados elaboraban, evaluaban y comunicaban planes de mercadotecnia. Todos eran líderes comerciales y se encargaban, además, de la organización y del control de la mercadotecnia. Otra entrevistada destacó: "para la ejecución de la organización de mercadotecnia uso los conocimientos de las temáticas gestión por procesos y procesos archivísticos, las cuales me han ayudado mucho para esto".

Con respecto a la pregunta sobre la medida en la que los conocimientos adquiridos en la carrera Bibliotecología y Ciencias de la Información han contribuido al desarrollo de las actividades mercadotécnicas, la figura 3 representa los elementos de mayor coincidencia.

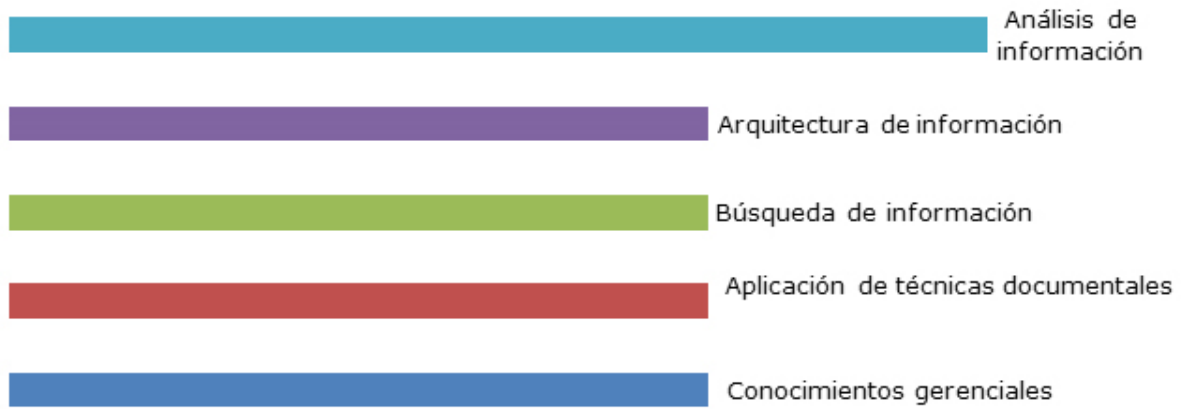

**Fig. 3.** Criterios de coincidencia en torno a los conocimientos obtenidos en la carrera que son de mayor uso en el desempeño de mercadotecnia.

244

Una entrevistada definió de forma específica: "entre los conocimientos de la carrera que me han ayudado se encuentran: la gestión por procesos, para dar seguimiento a las acciones y a los resultados. La archivística, para facilitarme el almacenamiento eficiente de los informes que genero cada día, y la gestión de información, para elaborar tales informes y presentarlos de una manera potable al consejo de dirección".

A través de los resultados obtenidos se pudo apreciar la coincidencia de los criterios en torno a los conocimientos obtenidos en la etapa académica. Mediante el procesamiento de las entrevistas se pudo observar que entre los conocimientos, habilidades y atributos personales que más se destacaron sobre la actividad de mercado, se encontraban los siguientes:

- *Conocimientos:* tratamiento de la información, gerencia de la información, investigación y tecnologías de la información.
- *Habilidades:* manejo de fuentes de información, toma de decisiones y solución de problemas basados en información, trabajo en equipo, gestión de información, de documentos y del conocimiento, y dominio de procesos asociados a la información.
- *Atributos personales:* cultura informacional, calidad en el ejercicio de la profesión, y motivación y compromiso hacia el ejercicio profesional.

Resulta curioso que la mayoría de los entrevistados mencionaba a las encuestas y entrevistas entre las herramientas que utilizaban, pero no explicitaron a la asignatura Metodología de la Investigación, la cual forma bases para esta manera de recopilar la información.

Los resultados del análisis de contenido aplicado al plan de estudios "D" mostraron que el documento cuenta con las materias clave referidas, las cuales están actualizadas y reformadas acorde con las exigencias del desarrollo actual de la profesión en general y de la mercadotecnia en lo específico. Los contenidos de la metodología de la investigación fueron potenciados con la inserción de una asignatura que permite una mayor práctica de los métodos y técnicas de investigación. Se reformularon los objetivos y contenidos de las asignaturas: Arquitectura de Información, Diseño de Proyectos, Mercadotecnia en Organizaciones de Información y Principios de Gestión de Sistemas y Organizaciones de Información. Además, se crearon otras nuevas como Comportamiento Humano en el Entorno Informacional y Taller de Comunicación Interpersonal, las cuales oportunamente tributan al trabajo de gestión mercadotécnica. Igualmente, con la posibilidad que brinda este plan de estudios de generar asignaturas teniendo en cuenta la necesidad social, laboral y contextual, se introducen otras asignaturas de carácter electivo y optativo que, de igual forma, inciden en el desarrollo del papel de los profesionales de la información como gestores de mercadotecnia.

## CONSIDERACIONES FINALES

La mercadotecnia, como área importante de la administración organizacional, está condicionada de forma intensiva por el uso de la información para la toma, implementación y evaluación de decisiones, así como la solución de problemas. Esto conlleva que cada actividad, tanto genérica como específica, sea informacional para que pueda ser ejecutada, lo cual condiciona que el gestor de mercadotecnia, más allá de sus competencias de negocio, requiere de conocimientos y habilidades informativas para realizar la gestión documental y de información para la toma de decisiones, elaborar y ejecutar los proyectos de investigación del mercado, así como tratar con cada persona, grupo o fuerza del entorno organizacional.

Durante la elaboración del plan de estudios "D" se deben tener en cuenta las necesidades del desarrollo social de Cuba y los avances científicos de los campos disciplinares que conforman las Ciencias de la Información. La perspectiva de renovación, actualización y reforzamiento de las competencias del egresado para su desempeño en cualquier esfera que usa la información intensivamente, también aplica al caso de los requerimientos de la actividad mercadotécnica. Por supuesto, tales competencias deben afianzarse en el transcurso del ejercicio de la profesión, en dependencia de las funciones concretas; pero los conocimientos, habilidades y valores obtenidos en la carrera de Ciencias de la Información ofrece a sus egresados las posibilidades del desenvolvimiento exitoso en el área de la gestión de mercadotecnia en el contexto empresarial cubano.

## REFERENCIAS BIBLIOGRÁFICAS

Lic. *Carlos Luis González-Valiente.* Departamento de Informática y Gestión de la Información del Grupo Empresarial de la Industria Sidero Mecánica (GESIME). La Habana, Cuba. Correo electrónico: cvaliente@sime.cu; carlos.valiente@fcom.uh.cu